\documentclass[aps,pre,showpacs, twocolumn,floatfix,superscriptaddress]{revtex4-1}
\usepackage{graphicx,hyperref,bbm,amssymb,amsfonts,amsmath,times,subfigure,color,soul, epstopdf}  

\def\ii{{\rm i}}
\newcommand{\ket}[1]{|{#1}\rangle}
\newcommand{\bra}[1]{\langle {#1}|}

\newcommand{\Title}{Fractality in nonequilibrium steady states of quasiperiodic systems}
\newcommand{\AAH}{Aubry-Andr\'{e}-Harper}
\newcommand{\AAHs}{AAH}

\date{\today}

\begin{document}

\title{\Title}
\author{Vipin Kerala Varma}
\affiliation{Abdus Salam ICTP, Strada Costiera 11, 34151 Trieste, Italy}
\affiliation{Initiative for the Theoretical Sciences, The Graduate Center, CUNY, New York, NY 10016, USA}
\affiliation{Department of Engineering Science and Physics, College of Staten Island, CUNY, Staten Island, NY 10314, USA}
\affiliation{Department of Physics and Astronomy, University of Pittsburgh, Pittsburgh, PA 15260, USA}
\author{Cl\' elia de Mulatier}
\affiliation{Abdus Salam ICTP, Strada Costiera 11, 34151 Trieste, Italy}
\author{Marko \v Znidari\v c}
\affiliation{Physics Department, Faculty of Mathematics and Physics, University of Ljubljana, 1000 Ljubljana, Slovenia}
\begin{abstract}
We investigate the nonequilibrium response of quasiperiodic systems to boundary driving. In particular we focus on the \AAH{} model at its metal-insulator transition and the diagonal Fibonacci model.
We find that opening the system at the boundaries provides a viable experimental technique to probe its underlying fractality, which is reflected in the fractal spatial dependence of 
simple observables (such as magnetization) in the nonequilibrium steady state. 
We also find that the dynamics in the nonequilibrium steady state depends on the length of the chain chosen: generic length chains harbour qualitatively slower transport (different scaling exponent) 
than Fibonacci length chains, which is in turn slower than in the closed system.
We conjecture that such fractal nonequilibrium steady states should arise in generic driven critical systems that have fractal properties.
\end{abstract}

\maketitle

\section{Introduction}
Mathematical and aesthetic beauty of fractals captures the imagination of scientist and nonscientists alike. 
They can be abundantly observed in nature, be it in the shape of a coastline or a broccoli flower. 
Indeed the human eye is better evolved to recognize and process fractal patterns rather than straight lines \cite{Taylor}.
Fractals can be found even in wavefunctions describing coherent quantum systems. 
Namely, it is well established that quantum systems at phase transition points can display fractal eigenfunctions. A paradigmatic example is the Anderson localization transition \cite{Anderson:58}: 
Below the transition point eigenfunctions are extended, above they are localized, while at the transition point their dimensionality is in-between, namely displaying noninteger fractal dimension \cite{Mirlin}. 
Despite that experimentally detecting fractality in quantum systems has been successfully demonstrated only rather recently from its local density of states~\cite{Richardella10}. 
The theoretical characterization of system fractality in closed systems is usually done either directly in terms of its wavefunction (eigenfunction) properties $-$ which, however, are notoriously difficult to measure experimentally $-$,
or through wavepacket spreading in unitary dynamics \cite{Abe, Huckenstein, Ohtsuki, Piechon}, or through return probability \cite{KravtsovCuevas}. 

\begin{figure}[bp!]
\centerline{\includegraphics[width=3.3in]{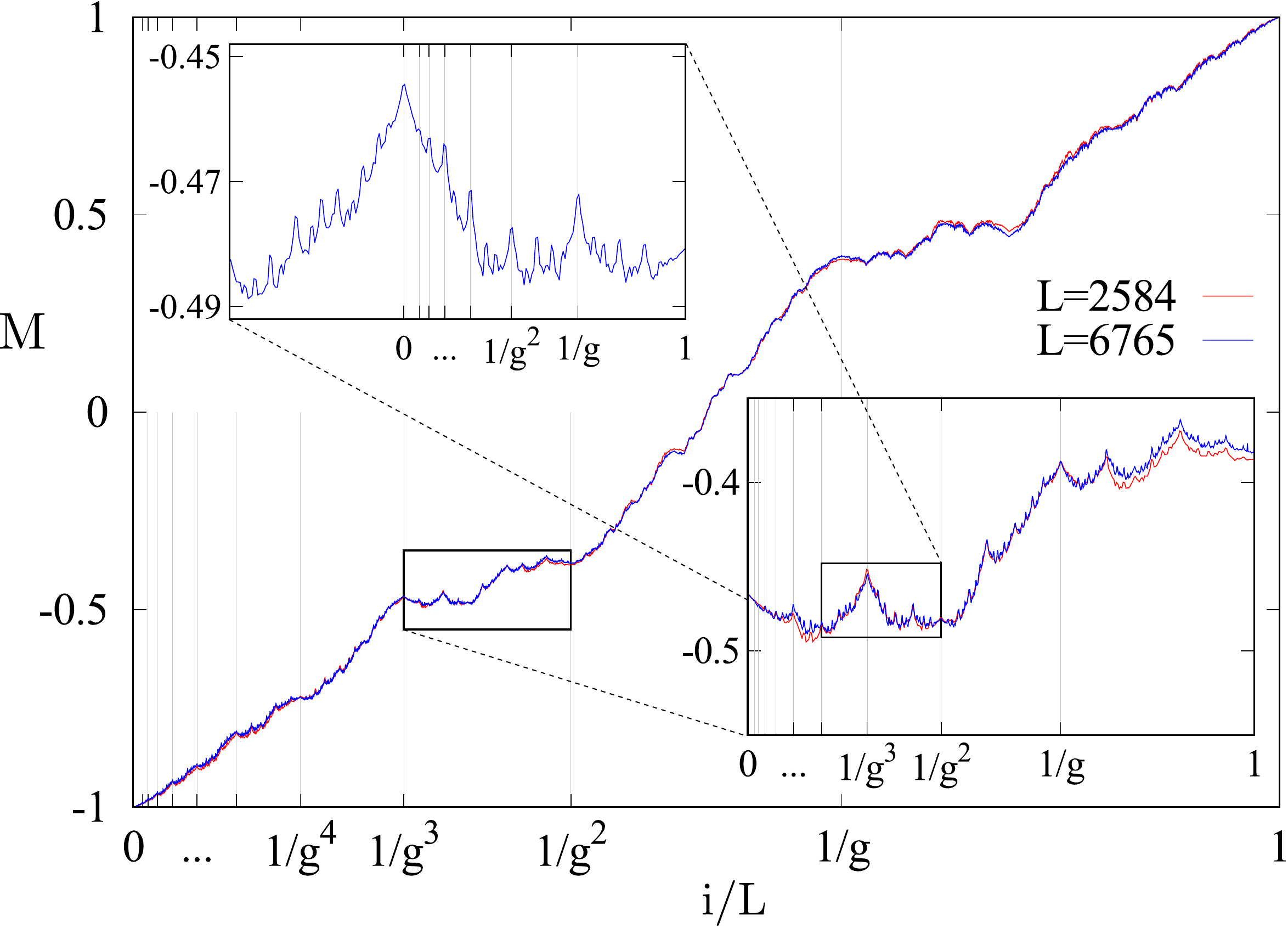}}
\vspace{-0.35cm}
\caption{(Color online) NESS magnetization profile in the \AAH{} model at criticality for Fibonacci chain lengths $L$. 
Each profile was obtained by averaging over $10^3$ $\phi$ values.
As one zooms in, finer and finer details are revealed with peaks located at fractions of the inverse golden ratio after rescaling along the $x$-axis~\cite{Supp}. 
Fractal dimension is $D_{\textrm{f}} \approx 1.11$ \cite{Supp}. 
The curves lie within the standard deviations of each other.
}
\label{fig:AAHprofileFib}
\end{figure}

In this article we show how such eigenfunction properties of a closed (Hamiltonian) system can also be brought to light in a rather transparent fashion if one couples the system at its boundaries to an external 
bath, thereby inducing a nonequilibrium steady state (NESS). We show that such a NESS displays fractal spatial dependence of generic observables, 
e.g., of particle density (Fig.\ref{fig:AAHprofileFib}), which is a quantity routinely accessible by today's experiments. The fractality that we reveal is not just fractal dependence on a system's parameter, 
like e.g. the dependence of transport coefficients in some classical~\cite{Klages} or quantum systems~\cite{Prosen}, but truly a fractal spatial property in a single-shot NESS. We demonstrate our findings on models with quasiperiodic potentials. 
The study of these systems have gathered significant traction vis-\`a-vis their localization and transport properties in cold atom experiments 
and photonic waveguide set-ups \cite{BlochLahini}. Experimentally realizing them is also considerably easier compared to the legacy disordered models introduced by Anderson \cite{Anderson:58}: a 
superposition of two incommensurate wavelengths readily creates such a quasidisordered potential. They are also of theoretical interest because they can display localization transition even in one dimension, a property expected to appear for disordered systems only in higher dimensions, which, however, are harder to treat analytically or numerically.

While localization properties of such non-interacting models have been theoretically investigated in Hamiltonian settings \cite{AubryAndre, Sutherland, AAFibonacciTransport, Abe, Piechon}, it is not known what their properties are upon external coupling, which is what necessarily happens in any experiment. We show that, besides the fractality of local NESS expectations, the fractal dimension and, even more interestingly, the scaling exponent of anomalous transport depends on the system's length: 
the number theoretic properties of the irrational number $g$ characterizing the incommensurability causes the system to distinguish between two classes of integers. Thus our finding, yet again, illustrates the richness of nonequilibrium quantum physics.

We also derive an exact connection between return probability of a simple non-Hermitian ``Hamiltonian'' and NESS transport. While rigorously defining transport properties in a Hamiltonian system can be quite tricky (e.g., exponentially decaying eigenfunctions do not necessarily imply localization~\cite{Rio95}), in an open system it is straightforward -- one simply studies the scaling of the current -- but one has to instead deal with non-Hermitian operators.

\section{Model and method}
The Hamiltonian of a quasidisordered spin chain of size $L$ is 
\begin{equation}
H=\sum_{i=1}^{L-1} \sigma_{i}^{x}\sigma_{i+1}^{x}+\sigma_{i}^{y}\sigma_{i+1}^{y}+ \sum_{i=1}^{L}h^{\phantom{z}}_i \sigma_i^z,
\end{equation}
where $\sigma_i^\alpha$ are Pauli matrices, and $h_i$ is the quasidisorder given by (i) $h \cos{(2\pi g i + \phi)}$ for the \AAH{} (\AAHs) model, with 
$g=\frac{\sqrt{5}+1}{2}$ the golden ratio, and $\phi$ an arbitrary global phase that is averaged over;  
(ii) $\pm h$ arranged in a Fibonacci sequence (see \cite{Supp} for details) for the diagonal Fibonacci model. 
The above Hamiltonian may be Jordan-Wigner transformed to a quadratic fermionic system \cite{Supp}, so that an exact solution of the steady state is possible (see below).

The \AAHs{} model is a paradigmatic model for studying delocalization-localization transitions in one-dimensional systems, with the transition at $h=h_c=2.0$ \cite{AubryAndre}: (i) for $h<h_c$ the spin transport is ballistic, (ii) for $h>h_c$ the system is localized, 
and there is no transport; 
(iii) at critical point $h=h_c$ eigenfunctions are neither extended nor localized but fractal \cite{Simon82, Ostlund}, and 
the transport is expected to be diffusive \cite{Guarneri}, or very close to diffusive \cite{Abe, Piechon}, due to the multifractality and singular continuous nature of the spectrum.

In the related Fibonacci model the spectrum however is \textit{always} composed of critical states \cite{Sutherland}; 
that is, there is no localization-delocalization transition like in the \AAHs{}, and this 1D quasidisordered model fully escapes localization (except when $h \rightarrow \infty$). 
Moreover the spin transport here for any finite $h$ is neither ballistic nor diffusive but shows a smooth variation of the transport between ballistic and no transport as a function of $h$ \cite{Abe}.

We use the Lindblad master equation~\cite{Lindblad} describing Markovian evolution of the system's density matrix,
\begin{equation}
\frac{{\rm d}\rho}{{\rm d}t}=\ii [ \rho,H ]+ \sum_{k=1}^4 \left( [ L_k \rho,L_k^\dagger ]+[ L_k,\rho L_k^{\dagger} ] \right) := \mathcal{L}(\rho),
\label{eq:Lin}
\end{equation}
where Lindblad operators $L_k$ drive magnetization through the boundaries, and are $L_1=\sqrt{\Gamma(1+\mu)}\,\sigma^+_1, L_2= \sqrt{\Gamma(1-\mu)}\, \sigma^-_1$ at the left end, 
and $L_3 =  \sqrt{\Gamma(1-\mu)}\,\sigma^+_L, L_4= \sqrt{\Gamma(1+\mu)}\, \sigma^-_L$ at the right end, with $\sigma^\pm_k=(\sigma^{x}_k \pm {\rm i}\, \sigma^{y}_k)/2$, 
and $\Gamma$ is the coupling strength to the bath; $\mathcal{L}$ is the Liouvillean characterizing the above master equation.
As long as there is a driving bias i.e. $\mu \ne 0$, a nonzero current is induced in the NESS $\rho_{\infty}$ uniquely given by $\rho_{\infty} = \lim_{t \rightarrow \infty} \textrm{exp}(\mathcal{L}t)\rho(t=0)$.
Because ${\cal L}$ is quadratic~\cite{Prosen10} all 2-point expectations in the NESS may be obtained by solving a Lyapunov equation $A C + C A^{\dagger} = P$, where the unknown 
elements of the complex Hermitian $L \times L$ correlation matrix $C$ specify the NESS~\cite{MarkoHorvat}, whereas the $A$ and $P$ matrices are specified by the Hamiltonian and the couplings to the reservoirs, 
$A_{i,k}=\ii [H_0]_{i,k}+\Gamma \delta_{i,k}(\delta_{i,1}+\delta_{i,L})$, 
see \cite{Supp} for technical details. Without loss of generality we set $\mu=\Gamma=1$. Using efficient numerical techniques we can study very large systems up to $L=32768$, 
which is crucial to correctly reveal both fractality and the asymptotic transport type.

\begin{figure}[bp!]
\centerline{\includegraphics[width=3.3in]{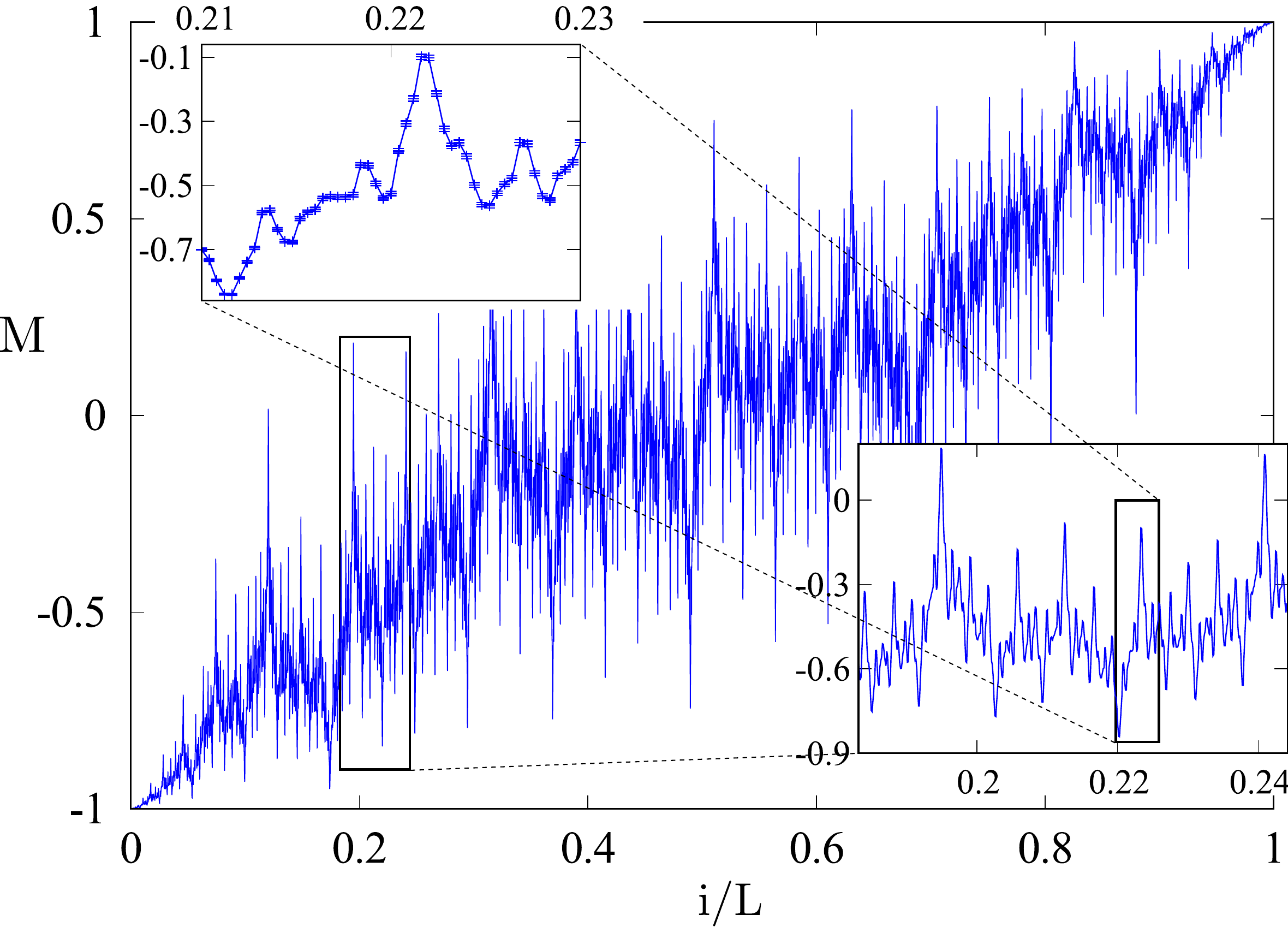}}
\vspace{-0.5cm}
\caption{(Color online) Magnetization profile in the \AAHs{} model at criticality for generic non-Fibonacci chain length $L=8192$. 
Fractal dimension is $D_{\textrm{f}} \approx 1.60$ \cite{Supp}, larger than in the case with Fibonacci lengths shown in Fig.~\ref{fig:AAHprofileFib}. 
Note that strong oscillations are not noise; in the inset, where we show individual points, the error bars (after averaging over $1000$ random phases) are of the same size as points.}
\label{fig:AAHprofile8192}
\end{figure}

\section{Nonequilibrium steady state}
We focus on expectation values, in the NESS, of the magnetization profile given by $M\equiv C_{i,i}=-\langle \sigma_i^z\rangle$
and the spin current given by $j \equiv \langle 2(\sigma_i^x \sigma_{i+1}^y - \sigma_i^y\sigma_{i+1}^x) \rangle = 4\textrm{Im}\{C_{i,i+1}\}$ at lattice site $i$ 
(by current conservation the latter is independent of $i$).

\subsection{Magnetization and fractality} 
In the steady state the magnetization profile is a particularly simple observable to measure, for instance by scanning tunneling electron 
microscope \cite{Richardella10} in solid state or spectroscopy \cite{BlochSpectroscopy} in cold atom set ups.
In addition the shape of the profile is an indicator of the rate of transport in the NESS \cite{ZnidaricVarma}: continuously ranging from a flat profile indicating ballistic, 
to a single-step function~\cite{Monthus17} indicating localization, with convex or concave profiles indicating subdiffusive or superdiffusive transport respectively. 

Let us focus on the critical \AAHs{} model, and let us first observe the magnetization profile for a Fibonacci length chain $L=6765=F_{20}$ displayed in Fig.~\ref{fig:AAHprofileFib} (where $F_n$ is the $n$-th element of the Fibonacci sequence).
Firstly we find that while the gross shape of the NESS magnetization profile conforms to that of subdiffusive behaviour rather than diffusive (for which the profile would have simply been a linear line) 
there are fine features atop this: there are valleys and peaks which do not disappear upon phase averaging, and whose number increase at smaller length scales as the system size is increased. 
Indeed as we keep zooming in finer features start appearing. In addition to this the structured features are prominently present at powers of 
the inverse golden ratio within the appropriately chosen window. Both these points already suggest the presence of fractality, which is further vindicated by the fractal box counting dimension $D_{\textrm{f}}$ calculated in Ref.~\cite{Supp}. 
We find that this dimension is not an integer but rather $D_{\textrm{f}} \approx 1.11$. 
However, away from criticality, say in the ballistic phase, we find that the box-counting dimension becomes 1 for Fibonacci $L$~\cite{Supp}.
This clearly shows that NESS can be a good probe of fractality present in critical systems. A similar result holds for the Fibonacci model as well \cite{Supp}, which is always critical.

Furthermore, if we study generic non-Fibonacci length chains, the underlying fractality is reflected in the profile both at and away from criticality. 
This is in line with fractal properties expected for quasiperiodic systems~\cite{Jitomirskaya}.
In Fig.~\ref{fig:AAHprofile8192} we show the profile for a chain of length $L=8192$. Clearly there are stronger oscillations and fluctuations compared to the Fibonacci length chain shown in 
Fig.~\ref{fig:AAHprofileFib} reflected in larger box-counting dimension $D_{\textrm{f}} \approx 1.6$.
Moreover there is a universality amongst the various system sizes even for non-Fibonacci lengths: $L$ and $\tilde{L} = L/g^n$ have overlapping magnetization patterns, 
despite the strong oscillations, and hence the same fractal dimension (for sufficiently large $L$) \cite{Supp}. This further vindicates our point that these oscillations are not noise.

Therefore, comparing the results of the systems of two classes of lengths (Fibonacci and non-Fibonacci lengths), 
we find that for $L=F_n$, real space resonances set in that causes destructive interference of these stronger oscillations, thereby lowering the 
$D_{\textrm{f}}$ (and increasing transport rate as we will see shortly).
This is found to be true even if we start moving closer towards a Fibonacci number.
The source of these resonances will ultimately be connected to the spectrum of $A$ because the latter can be directly related to the NESS \cite{Supp}.
The onset of fractality in the magnetization profiles, and in fact in any other observable~\cite{Supp}, and its sensitive dependence on the chain length is the first main finding of our work.

\begin{figure}[ttp!]
\centerline{\includegraphics[width=3.3in]{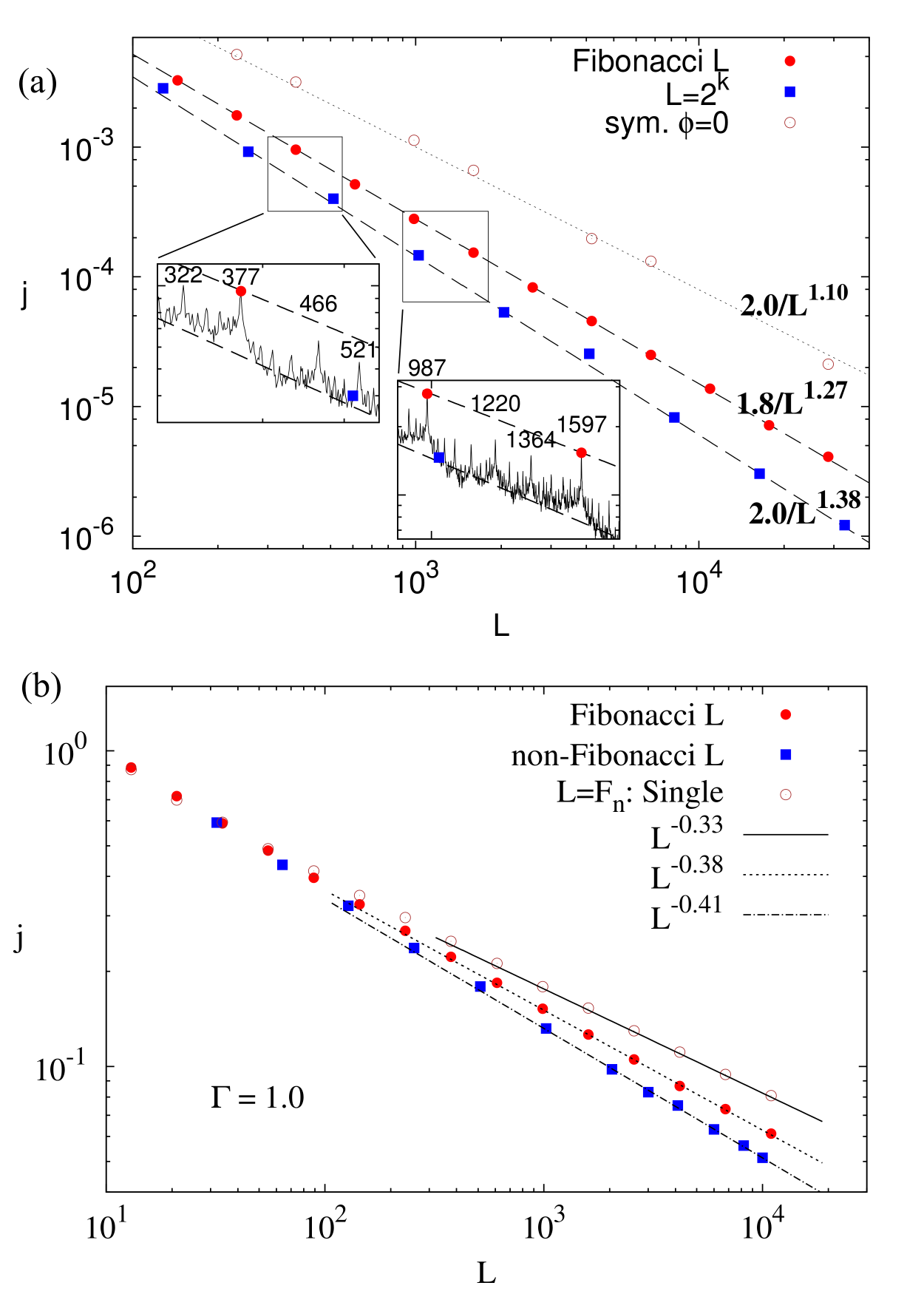}}
\vspace{-0.35cm}
\caption{(Color online) NESS current at criticality. Average current is shown, except in the single-shot cases (open symbols).
Top panel: The \AAHs{} model where the scaling is subdiffusive, and depending on the length, Fibonacci vs. generic vs. single shot symmetric quasidisorder, the scaling exponent $\gamma$ is different. 
The insets show resonances in the NESS current at the Fibonacci length (red circles); few of the many ``satellite'' resonances at $F_n\pm F_m$ are also highlighted 
(this is equivalent to subdividing the intervals as in Fig.~\ref{fig:AAHprofileFib}), which are not noise but each of which have a sequence and scaling associated with them \cite{Supp}.
The system with the symmetric quasiperiodic potential is almost diffusive, as in the closed system dynamics. 
Although we present results for $\Gamma = 1.0$, we have checked that a weaker coupling $\Gamma=0.1$ shows similar qualitative behaviour.
Bottom panel: The Fibonacci model with $h=0.5$ showing similar qualitative dependence on system length and quasidisorder realization, is superdiffusive in all three cases. 
In both panels $L=4096$ deviates from the generic scaling because it is quite close to a secondary resonance i.e. $4096 \approx F_{19} - F_{11} = 4181 - 89$.
}
\label{fig:jL}
\end{figure}

\subsection{Current scaling and dynamics} 
One of the cleanest manifestations of nonequilibrium dynamics is transport. The rate of transport is measured by power-law scaling of current $j$ with system size. 
Assuming a phenomenological transport law $j = -D\nabla \sigma^z$, where $D$ is the diffusion constant, depending on the current scaling $j \sim 1/L^{\gamma}$, we have (i) $\gamma = 1$ for diffusion, (ii) $\gamma < 1$ for superdiffusion, (iii) $\gamma > 1$ for subdiffusion, 
(iv) $\gamma \rightarrow \infty$ in the localized phase, and (v) $\gamma = 0$ for ballistic transport.

In Fig.~\ref{fig:jL} we show the current $j$ as a function of $L$, in the critical \AAHs{} model in the top panel and in the diagonal Fibonacci model with $h=0.5$ in the bottom panel.
We see that three distinct types of scalings $j \sim 1/L^{\gamma}$ exist in both critical systems: 
$\gamma_{\textrm{Fibo.}}, \gamma_{\textrm{gen.}}, \gamma_{\textrm{sing.}}$ corresponding to Fibonacci length chains, generic length chains, and single special realizations of the quasidisorder, respectively.
Fibonacci length systems harbour slightly faster dynamics than generic length systems, and the single-shot quasidisorder realizations in turn faster than both (in \AAHs{} model this special single-shot 
realization is chosen such that the quasiperiodic potential is symmetric with respect to the central site of the chain, whereas for the Fibonacci model this sample is chosen as the one starting from the 
beginning of the Fibonacci sequence described in Ref. \cite{Supp}).

We therefore find quite odd behavior, unobserved previously, that the nature of transport in a NESS depends on the number theoretic properties of system's length. 
This is the second main finding of our work. 
Note that the transition point between ballistic ($h<2$) and localized phase ($h>2$) still occurs at $h=2$, the same as in the closed system: marginally slightly away from $h_c$ the scaling considerably changes \cite{Supp}.
The differing transport rate in the non-Fibonacci $L$ arise quite likely due to the stronger ``oscillations'' seen in its magnetization profile 
(Fig.~\ref{fig:AAHprofile8192}), which in turn give rise to more oscillations in the local resistivities.
Note that in order to clearly distinguish the different regimes one needs systems of several thousand sites (Fig.~\ref{fig:jL}); for instance, in the \AAHs{} model the resonances 
(see the insets) have a width of $\approx 10$, and therefore they begin to overlap with neighboring satellites for $L \lesssim 10^2$. 

\subsection{Spectral properties}
As outlined in Ref.~\cite{Supp}, one can directly express the NESS current in terms of a non-Hermitian $L \times L$ matrix $A$. Defining a non-unitary propagator $U(t) \equiv {\rm e}^{-At}$, we can write
\begin{equation}
j=8\left(1-2\int_0^\infty{\!\!\!\!p_1(t){\rm d}t}\right),
\label{eq:pj}
\end{equation}
where $p_1\equiv |U_{1,1}(t)|^2$ is transition probability from site $1$ to site $1$ (i.e., return probability). While $p_1(t)$ will in general have a power-law asymptotic decay due to singular density of eigenvalues of $A$~\cite{Supp}, 
we have not been able to find a simple connection between $\gamma$ and the spectral properties of $A$ characterized by the exponent $\kappa$ \cite{Ossipov01, Supp}; 
elucidating a scaling relation, if any, among the two exponents remains an interesting open problem, especially since $\gamma$ depends on the number theoretic properties of the chain length. 
The absence of such a relation, however, without invoking an additional \textit{independent} critical exponent 
will highlight the nontrivial effects of fractality \cite{Supp}.

\section{Closed system dynamics} 
Having investigated the response of these systems to Lindblad boundary drives, let us study the dynamics of wavepacket evolution under Hamiltonian dynamics, 
focussing on the \AAHs{} model at criticality. In contrast to the open system there is no system size dependence here because 
the wavepacket spreading does not feel the system length until the boundary is hit.
We initialize the system with a delta function at the central site and compute the mean-squared displacement $\Delta x^2 (t) = \sum_x \left[x - (L+1)/2\right]^2 |\langle x | \psi(t) \rangle|^2$, 
where $|\psi(t) \rangle = \textrm{exp}(-\textrm{i}t H) |\psi(0)\rangle$ is the unitarily evolved initial state with the \AAHs{} Hamiltonian. 
A scaling fit $\Delta x^2 \sim t^{2\beta}$ is employed to discern the rate of transport: $\beta = 1$ for ballistic, $\beta = 0$ for no transport.

\begin{figure}[tp!]
\centerline{\includegraphics[width=3.3in]{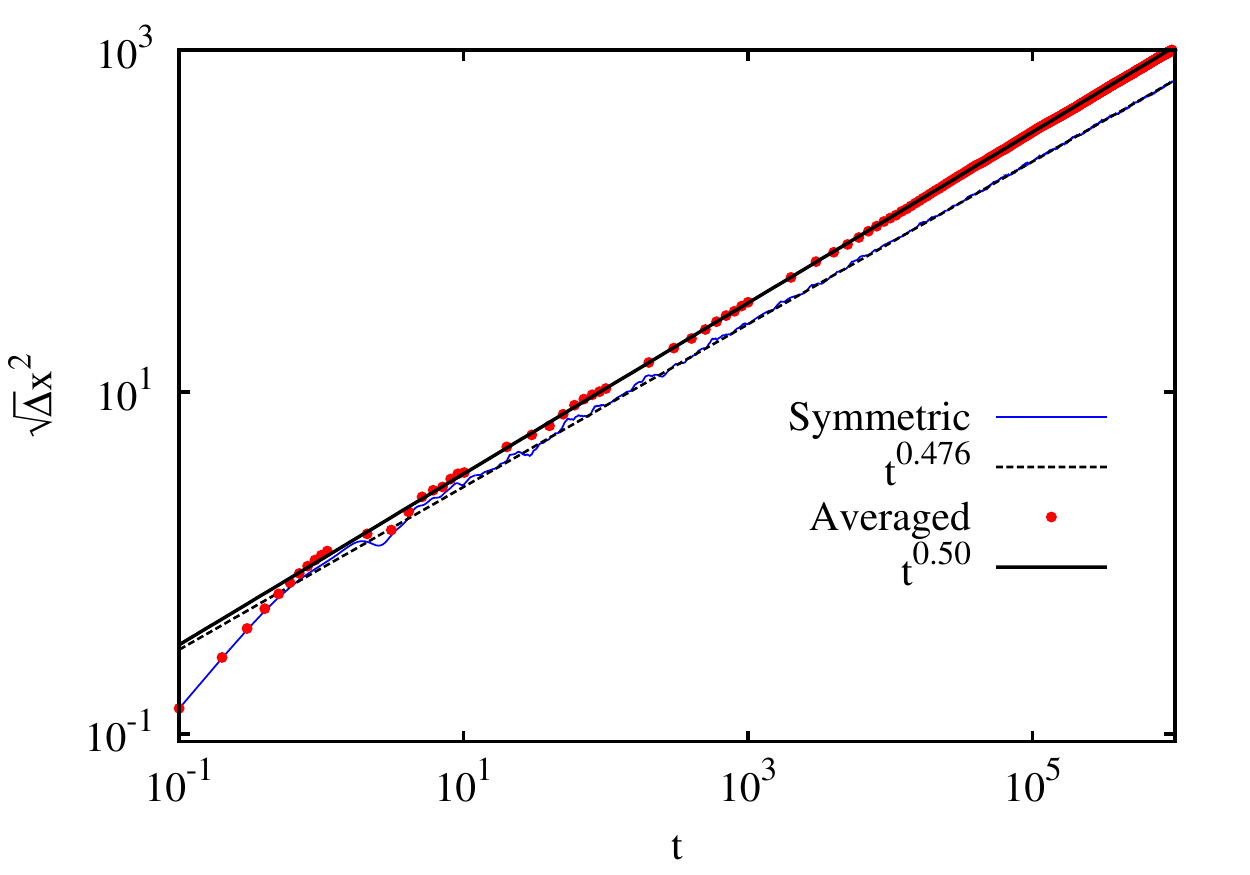}}
\vspace{-0.25cm}
\caption{(Color online) Root mean-squared displacement in wavepacket spreading with Hamiltonian dynamics in critical \AAHs{} Hamiltonian with free boundaries, $L=8001$. 
For symmetric realization of the quasidisorder (blue full line), 
a nice fit (dashed black line, indistinguishable from data) to subdiffusive spreading is observed whereas upon phase averaging (red dots, $\approx 500$ samples) normal diffusion is restored.} 
\label{fig:AA_closed}
\end{figure}

Our findings are two-fold: firstly we show that if we choose a symmetrized version of the quasiperiodic potential such that it is symmetric with respect to the central site ($\phi=0$), then we recover subdiffusive behaviour, consistent with earlier studies \cite{Abe, Piechon}; note that there is no averaging over the global phase $\phi$, 
which destroys this symmetry as soon as $\phi \neq 0$. This result is shown as the thin line in Fig.~\ref{fig:AA_closed} with the fit plotted as dashed line: the two are virtually indistinguishable. 
Although subdiffusive, it is quite close to being diffusive with $\beta = 0.476$; the value is remarkably close to the effective dynamical exponent $\beta_{\textrm{eff.}} = 1/(1 + \gamma)$ \cite{ZnidaricVarma} 
obtained for the open system dynamics with the symmetric potential i.e. $\beta_{\textrm{eff.}} = 1/(1 + \gamma_{\rm sing}) = 1/(1+1.10) \approx 0.476$. 
However, upon phase averaging this symmetry of the potential is broken and we restore normal diffusion, $\beta=1/2$, as indicated in the Fig.~\ref{fig:AA_closed} by the upper curve and data points. 
We note that in this case closed system $\beta$ does not agree with the NESS one via the relation $1/(1+\gamma)$. 
This failure, as that between $\kappa$ and $\gamma$, is quite likely symptomatic of multifractality in the system where the assumption of a single-exponent breaks down.
As opposed to the open system, in Hamiltonian formulation transport gets faster once the potential symmetry is broken, whereas in the NESS the symmetric case was the fastest. 
This is the third main finding of our work.

\section{Conclusions} 
In one-dimensional noninteracting systems in the presence of correlations in the potential, criticality and multifractality can be induced in the spectrum; 
the simplicity of these systems offers the scope of experimentally probing for such nontrivial physics through cold-atom or solid state wire set-ups, explaining recent high interest in quasiperiodic systems. 

Focusing on the critical \AAH{} model and the diagonal Fibonacci model, we demonstrate that in a nonequilibrium setting induced by boundary driving the underlying fractality of the system 
is made apparent in the nonequilibrium steady state expectation values of simple observables. 
In particular we show that the nonequilibrium steady state magnetization profile is fractal, and that the transport rate of spins depends sensitively 
on the length of the chain, in particular on whether the length is an integer related to the incommensurability of the quasiperiodic potential or not. 
The spin transport rates in the open setting seem in general different (slower) than in the critical closed system Hamiltonian dynamics.

We posit that these new rich emergent features in nonequilibrium physics are a consequence of underlying fractality, providing us a new and perhaps simpler probe for these phenomena in an open setting. 
We also rigorously relate transport to spectral properties of a simple non-Hermitian deformation of an almost Mathieu operator, whose Hermitian version displays very rich mathematical structure~\cite{Fields}.

\textit{Note added}: After completion of our work Ref. \cite{Kulkarni} appeared, which studies nonequilibrium transport in the critical \AAHs{} model at generic lengths, without reference to the fractal features 
in the system or the current resonances that show up at nongeneric lengths.

M.\v Z. acknowledges Grant No. J1-7279 from the Slovenian Research Agency (ARRS). This work has been supported by the ERC grant OMNES.

\section*{Appendix}

\subsection{NESS solution}

\subsubsection*{Steady-state equations}

Using Jordan-Wigner transformation \cite{Wigner} our models can be equivalently written in terms of fermionic operators $c^{\phantom{\dagger}}_j, c_j^\dagger$. 
To be precise we may employ the transformation \cite{Wigner} 
$\sigma^x_j = -(\sigma^z_1 \ldots \sigma^z_{j-1})(c^{\phantom{\dagger}}_j + c^{\dagger}_j)$, $\sigma^y_j = -\textrm{i}(\sigma^z_1 \ldots \sigma^z_{j-1})(c^{\phantom{\dagger}}_j - c^{\dagger}_j)$, 
and $\sigma^z_j = c^{\phantom{\dagger}}_j c^{\dagger}_j - c^{\dagger}_j c^{\phantom{\dagger}}_j$, to convert the spin-Hamiltonian to a quadratic fermion model.

With this the fermionic Hamiltonian reads \cite{MarkoHorvat}
\begin{equation}
 \label{eq: fermionicH}
 \mathcal{H}_{\textrm{f}} = \sum_j2\left(c^{\dagger}_j c^{\phantom{\dagger}}_{j+1} + c^{\dagger}_{j+1} c^{\phantom{\dagger}}_{j}\right) + \sum_j h_j(1 - 2n_j),
\end{equation}
where $n_j = c^{\dagger}_jc^{\phantom{\dagger}}_{j}$ is the counting operator.
This Hamiltonian, as well as the complete Liouvillean ${\cal L}$ 
(because the Lindblad operators $L_j$ are linear in $c_j$), is quadratic in $c_j$. In analogy with quadratic Hamiltonians one can therefore fully diagonalize ${\cal L}$ by finding non-interacting decay modes, and in particular, write down a closed set of equations for the steady-state expectation values of all 2-point fermionic observables~\cite{Prosen10}.

Let us expatiate on the technicalities. The procedure we outline in fact works even for the model with e.g. dephasing dissipation, which renders ${\cal L}$ non-quadratic, though still enabling one to write a 
closed set of equations for all 2-point NESS expectations. 
We shall follow the notation in Ref.~\cite{MarkoHorvat} where such more general case of a disordered XX model with dephasing has been studied. 
Writing the NESS as 
\begin{equation}
 \label{eq: NESS}
 \rho_{\infty} = \cfrac{1}{2^L}[1 + \mu({\cal H} + {\cal B})] + \mathcal{O}(\mu^2)
\end{equation}
where ${\cal H} = \sum_{r=1}^L\sum_{j=1}^{L+1-r}h_j^{(r)}H_j^{(r)}$ and ${\cal B} = \sum_{r=2}^L\sum_{j=1}^{L+1-r}b_j^{(r)}B_j^{(r)}$, with  
$H_j^{(r+1)} (B_j^{(r+1)}) \equiv \sigma_j^x Z_{j+1}^{[r-1]}\sigma_{j+r}^{x(y)} \pm \sigma_j^y Z_{j+1}^{[r-1]}\sigma_{j+r}^{y(x)}$, for $r \geq 1$, and $Z_j^{[r]} \equiv \prod_{k=j}^{j+r-1}\sigma_k^z$, while $H^{(1)}_j=-\sigma_j^{z}$.
Note that the above solution is valid for \textit{any} value of $\mu$ because the higher order terms are all orthogonal to operators in ${\cal H}$ and ${\cal B}$, see Ref.~\cite{MarkoHorvat} for more details, where also the fermionic version is written out explicitly. Therefore, we need only solve for 
the unknown coefficients $h_j^{(r)}, b_j^{(r)}$, and thence the corresponding observable's expectation value in the NESS is known.  
Because these expectations values are trivially proportional to $\mu$ we set $\mu = \Gamma = 1$.

The $h_j^{(r)}, b_j^{(r)}$ variables, which are equal to the NESS expectation values of the corresponding operators $H_j^{(r)}$ and $B_j^{(r)}$, and thence the NESS solution, are obtained from solving a Lyapunov equation,
\begin{equation}
A C + C A^{\dagger} = P,
\label{eq:Lyap}
\end{equation} 
where the elements of the correlation matrix $C$ are specified as $C_{j,k} \equiv h_j^{(k-j+1)} + \textrm{i}\, b_j^{(k-j+1)}$ for $k>j$, $C_{j,j} = h_j^{(1)}$, with 
$C_{j,k} = C_{k,j}^{*}$ for $j>k$.  
In particular, the NESS magnetization profile at lattice site $j$ is given by $\langle \sigma_j^z \rangle = -C_{j,j}$ 
and the spin current by $\langle 2(\sigma_j^x \sigma_{j+1}^y - \sigma_j^y\sigma_{j+1}^x) \rangle = 4\, \mathrm{Im}C_{j,j+1}$, which is of course in the NESS independent of site $j$.
The $A$ and $P$ matrices are determined by the Hamiltonian and the driving: $A \equiv \textrm{i}(E - J) + \Gamma R$, with $E_{j,j} = h_j$ 
determined by the quasidisorder, $J_{j,j+1} = J_{j+1,j} = 1$ represents the hopping, $R_{1,1} = R_{L,L} = 1$, and $P_{1,1} = -P_{L,L} = -2\Gamma$. All unspecified matrix elements are zero.

\subsubsection*{Spectral connection}
\begin{figure}[bp!]
\centerline{\includegraphics[width=3.3in]{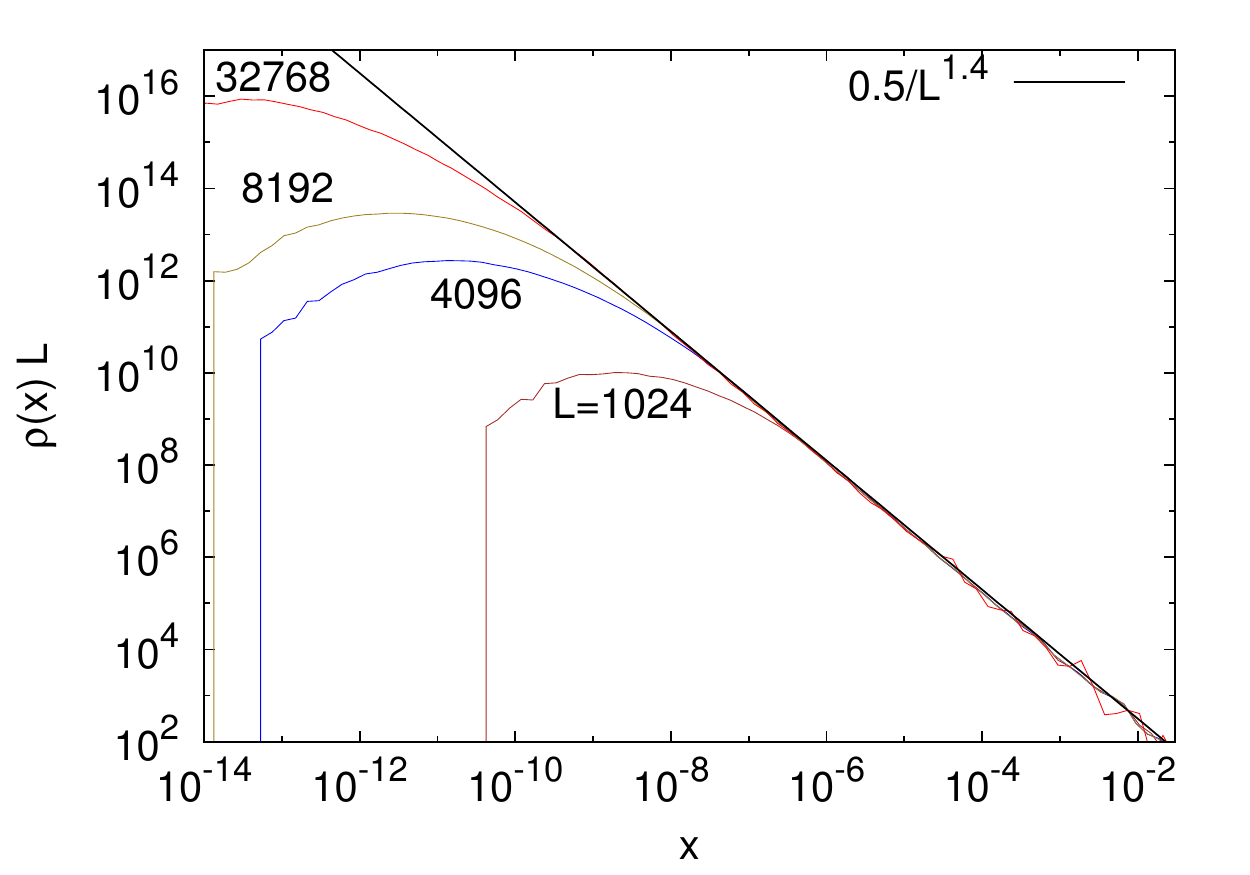}}
\vspace{-0.5cm}
\caption{(Color online) Distribution of the real parts $x$ of complex eigenvalues $\lambda_j$ of $A$. For Fibonacci lengths (data not shown) the power is also $\approx 1.40$.}
\label{fig:rhox}
\end{figure}
\begin{figure}[tp!]
\centerline{\includegraphics[width=3.1in]{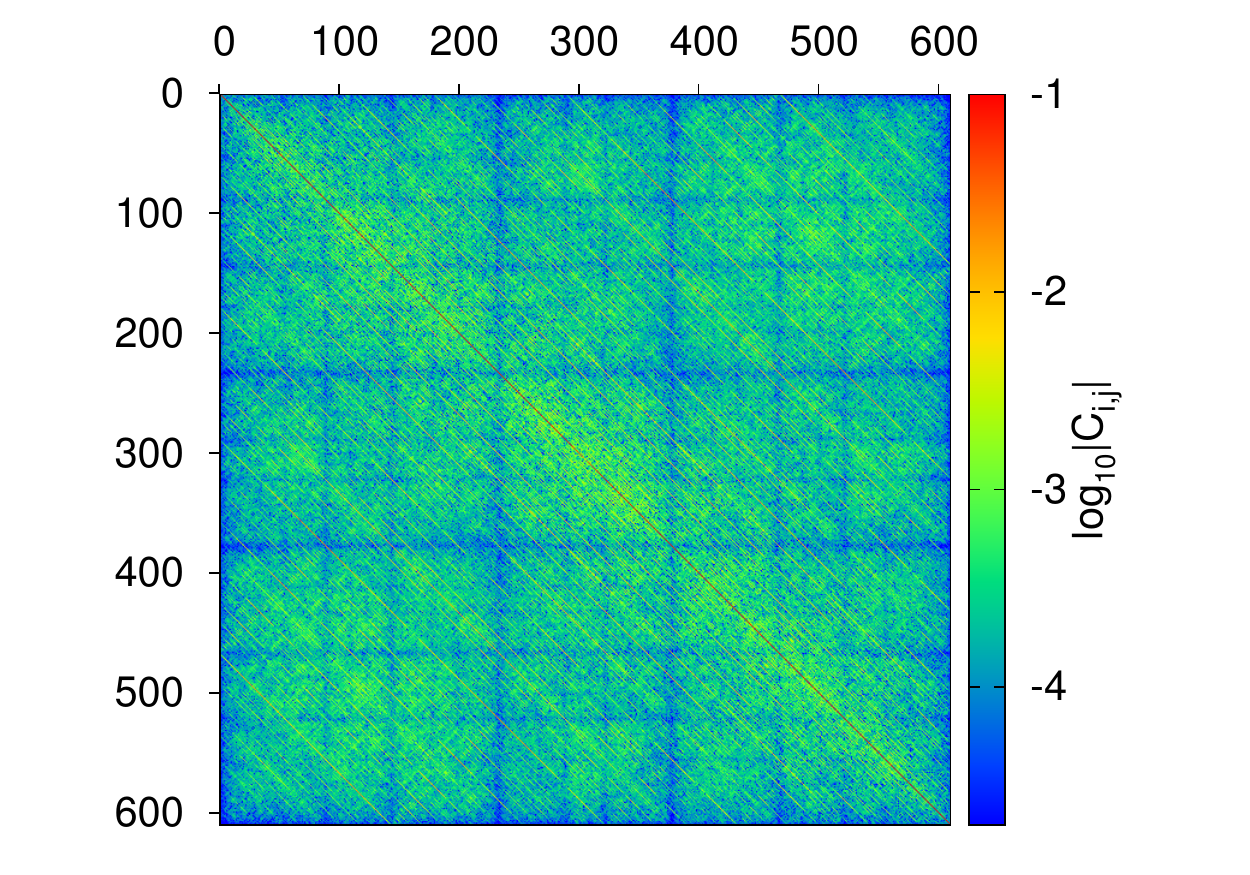}}
\vspace{-0.5cm}
\caption{(Color online) Two-point correlations in the NESS for the critical \AAH{} model with $L=610$ (Fibonacci length). We averaged over $1000$ random phases $\phi$. Minima in the correlations 
(vertical and horizontal ``blue'' lines occur for Fibonacci numbers $j$, and a self similarity is visually present here too as with the magnetization.}
\label{fig:C_n610}
\end{figure}
One can solve the Lyapunov equation (\ref{eq:Lyap}) by any standard linear algebra package, or, alternatively, one can express the solution in terms of spectral properties of the non-Hermitian matrix $A$, which we shall outline here.
Note that the matrix $A$ (setting $\Gamma=1$),
\begin{equation}
-\ii A=H_0 -\ii 
\begin{pmatrix}
1 & 0 & \cdots &  & 0\\
0 & 0 & \cdots &  & 0\\
\vdots & & \ddots & & \vdots \\
 & & \cdots & 0 & 0\\
0 &  & \cdots & 0 & 1
\end{pmatrix}
\label{eq:SA}
\end{equation}
is a sum of a Hermitian part $H_0\equiv E-J$, coming from a Hamiltonian of a single-particle disordered tight binding model, and an imaginary deformation given by driving $R$. It therefore represents the simplest non-Hermitian deformation of a random Schr\" odinger operator (of an almost Mathieu type) much studied in mathematics, see e.g. Ref.~\cite{Simon82, Simon2000}. Spectral properties of such a matrix should be an interesting future problem in itself.

Formally, the solution of the Lyapunov equation (\ref{eq:Lyap}) can be written~\cite{Lancaster70} as 
$C=\int_0^\infty {\rm e}^{-t A}P {\rm e}^{-t A^\dagger} {\rm d}t$.
Denoting a non-unitary ``propagator'' 
\begin{equation}
U(t) \equiv {\rm e}^{-t A}, 
\end{equation}
we can write $C_{p,k}=-4\int_0^\infty [U(t)]_{p,1} [U^\dagger(t)]_{1,k} {\rm d}t$ for $p>k$ (where due to symmetry it is enough to take only one of nonzero elements of $P$), and $C_{k,k}=2\int_0^\infty{([U(t)]_{k,L} [U^\dagger(t)]_{L,k}-[U(t)]_{k,1} [U^\dagger(t)]_{1,k}){\rm d}t}$. In our case $A$ seems always diagonalizable, 
so using the spectral decomposition $A=\sum_i \lambda_i \ket{\psi^{(Ri)}}\bra{\psi^{(Li)}}$, 
where it turns out that the left eigenvectors are just the complex conjugated right ones, $\psi^{(Lj)}_k=(\psi^{(Rj)}_k)^*$, one for instance gets the NESS current
\begin{equation}
j=-16\, {\rm Im} \left[ \sum_{i,k} \frac{1}{\lambda_i +\lambda_k^*} (\psi^{(Ri)}_1)^2 (\psi^{(Rk)}_1 \psi^{(Rk)}_2)^* \right],
\label{eq:SC}
\end{equation}
where everything is written in terms of the $k$-th components of the $i$-th right eigenvector $\psi^{(Ri)}_k$, which are, due to $\langle \psi^{(Lk)} | \psi^{(Ri)}\rangle =\delta_{ik}$, 
normalized such that $\sum_p \psi_p^{(Ri)} \psi_p^{(Rk)}=\delta_{ik}$. Provided one needs only the current, diagonalization of $A$ and using the above formula (\ref{eq:SC}) 
is the fastest way to compute it. The expression for magnetization is on the other hand
\begin{equation}
\langle \sigma_p^z \rangle=-1+4 \left[ \sum_{i,k} \frac{1}{\lambda_i +\lambda_k^*} (\psi^{(Ri)}_1 \psi_p^{(Ri)}) (\psi^{(Rk)}_1 \psi^{(Rk)}_p)^* \right].
\label{eq:Sz}
\end{equation}

While the eq.(\ref{eq:SC}) is useful for numerical computation, it is not very transparent. One can in fact get physically more revealing expression by observing that in the steady state the continuity equation at the chain edge~\cite{MarkoHorvat} is $j=4(1+C_{1,1})$. This leads to the NESS current
\begin{equation}
j=4\left(1-2\int_0^\infty{\!\!\!\!\!(p_1-p_L){\rm d}t}\right)=8\left(1-2\int_0^\infty{\!\!\!\!p_1 {\rm d}t}\right),
\label{eq:Sj}
\end{equation}
where $p_1\equiv |U_{1,1}(t)|^2$ is transition probability from site $1$ to site $1$ (i.e., return probability) and $p_L\equiv |U_{1,L}(t)|^2$ is transition probability from site $L$ to site $1$, both under nonunitary evolution with $A$. In the last equality we used $\int_0^\infty {(p_1+p_L){\rm d}t}=\frac{1}{2}$. 
Recall that $U_{k,i}(t)=\sum_p {\rm e}^{-\lambda_p t}\psi_k^{(Rp)} \psi_i^{(Rp)}$. If there were only dissipation 
and no dynamics, $p_{1,L}$ would decay exponentially in time, so that their integral would cancel $1$ in eq.(\ref{eq:Sj}), giving $j=0$. 
Therefore, their deviation from an exponential decay eventually determines the NESS current.

Let us show that although the eigenvalues of $A$ are pivotal in determining the scaling of $j$ with $L$, they are not sufficient. 
Return probability in the \AAH{} model has been studied in Ref.~\cite{Ossipov01}. Phenomenologically opening the system by adding an imaginary matrix element to the Hamiltonian, like in our rigorously derived matrix $A$ (\ref{eq:SA}), 
the authors find that the return probability decays as a power law with the power being in turn determined by the spectral properties of $A$. Writing the complex eigenvalues $\lambda_j \equiv x_j+\ii y_j$, 
one can use perturbation theory and argue that for small $R$ the real parts $x_j$ (``resonance widths'') will be proportional to the overlaps, $x_j \sim |\psi_1^{[j]}|^2$, where $\psi^{[j]}$ is the $j$-th eigenvector of $H_0$. Using a spectral expression for $p_1(t)$, averaging over fast oscillations, 
one gets $p_1(t) \asymp \sum_j x_j^2 {\rm e}^{-2x_j t} \approx L \int \rho(x)x^2 {\rm e}^{-2xt}{\rm d}x$, where $\rho(x)$ is a normalized distribution of the real parts of $\lambda_j$. 
This distribution has a power-law divergence for small $x$, which would then in turn, via Eq.(\ref{eq:Sj}), determine the scaling of current with $L$. 
We find, however, that the current scaling $j \sim 1/L^\gamma$ is more complex and can not be explained solely by the properties of $\rho(x)$. 
For instance, in Fig.~\ref{fig:rhox} we show the distribution $\rho(x)$, which for small $x$ indeed diverges as $L\rho(x) \sim 1/x^\kappa$, with $\kappa \approx 1.40$ though being the same for 
Fibonacci number $L$ and non-Fibonacci length systems, whereas the scaling exponent $\gamma$ on the other hand does depend on $L$. 
That the values of $\gamma$ cannot be inferred only in terms of eigenvalues of $A$ is expected in light of a sensitive dependence of $\gamma$ on the location of the driving. 
Such information can namely be only encoded in the eigenvectors of $A$ and so correlations between eigenvalues and eigenvector components in e.g. eqs. (\ref{eq:Sj}) or (\ref{eq:SC}) do matter. 

\subsection{Correlation matrix}
\begin{figure}[tp!]
\centerline{\includegraphics[width=3.1in]{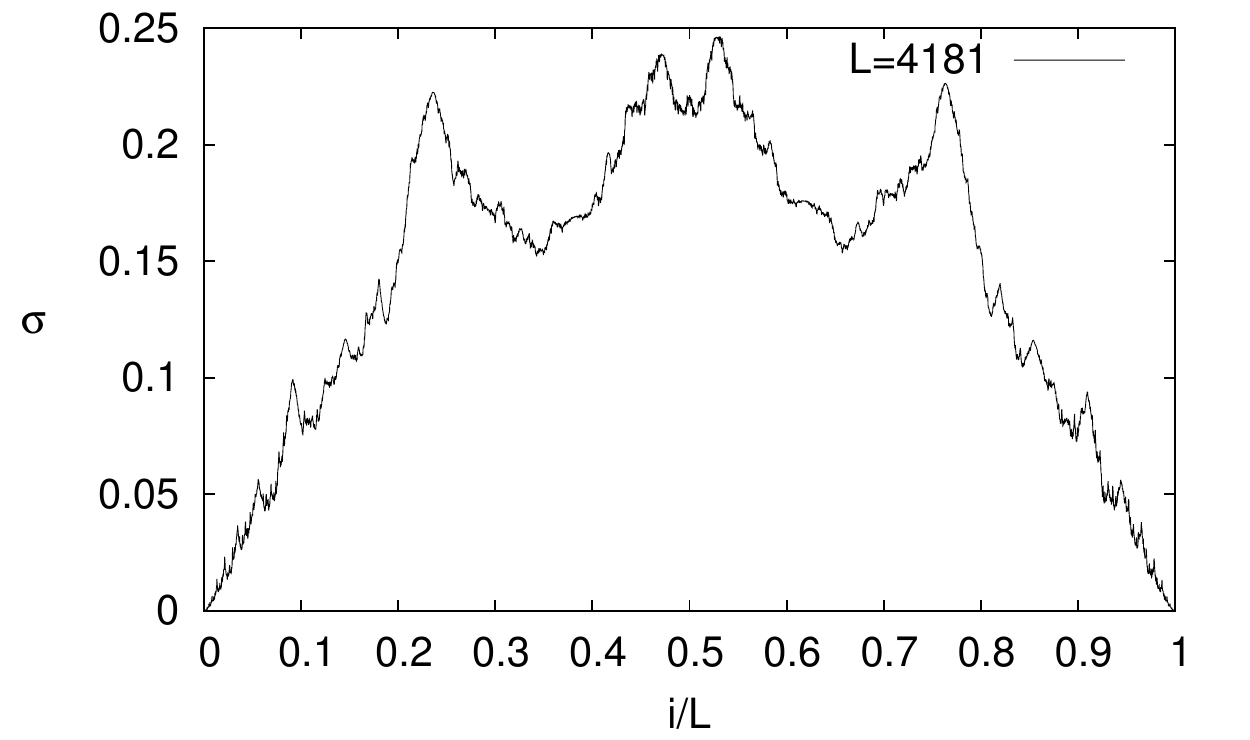}}
\vspace{-0.4cm}
\caption{Fractal dependence of standard deviation of magnetization $\langle \sigma_i^z \rangle$ in the \AAHs{} 
model on the rescaled spatial index. 
Fibonacci length $L=4181$ was chosen. 
}
\label{fig:AAHsigmaFib}
\end{figure}
\begin{figure*}[t]
\centerline{\includegraphics[width=6.5in]{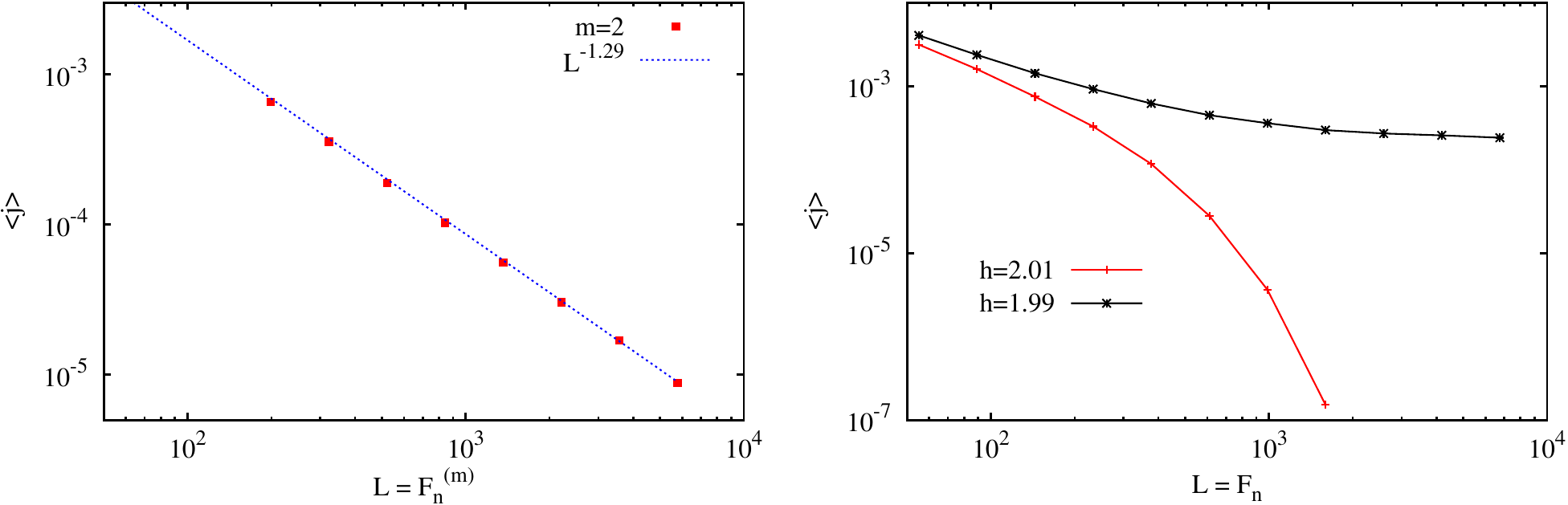}}
\vspace{-0.2cm}
\caption{(Color online) Nonprimary resonances and off-criticality. Left panel: Scaling of phase-averaged NESS current for $m=2$ Fibonacci length sequence (Eq. \eqref{eq: mFibo}) 
in the critical \AAH{} model. A nice power-law subdiffusive scaling with exponent 1.29 (very close, if not equal within statistical errors, to $m=1$ shown in main text) is obtained. 
This vindicates the point that the ``satellite'' resonances displayed in Fig. 3 of the main text are real in that they follow a scaling depending on the particular length sequence chosen, and that they 
are not noise.
Right panel: Scaling of NESS current slightly away from criticality. Immediately to the left of the closed system critical point $h_c = 2.0$, we find a saturation of the current setting in as the system size 
is increased, signalling ballistic transport of spin in the metallic phase. Whereas immediately to the right of $h_c$ we find an exponential decay of the current, characteristic of localization and absence of 
any transport.
All plots correspond to $\Gamma = 0.1$.
}
\label{fig:2ndFibo}
\end{figure*}
\begin{figure}[bp!]
\centerline{\includegraphics[width=3.3in]{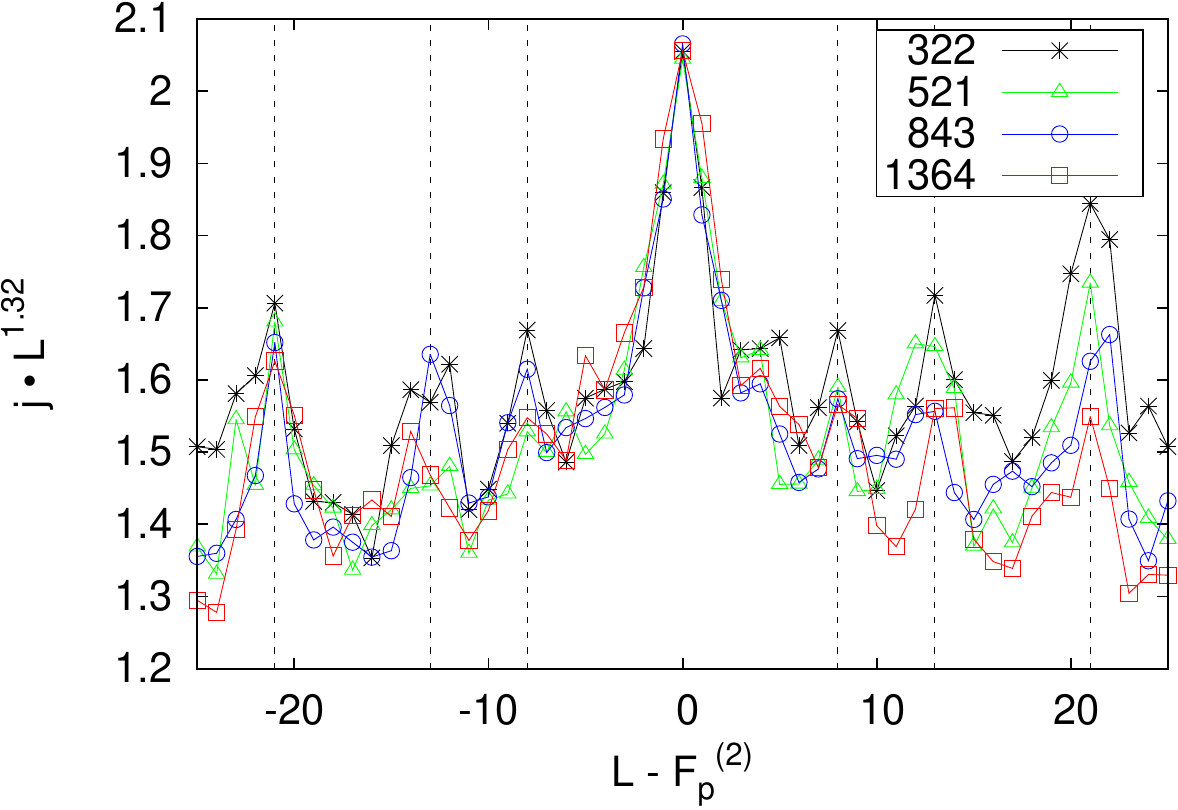}}
\vspace{-0.5cm}
\caption{(Color online) Stability of secondary resonances. Scaled NESS currents as a function of deviation from the secondary Fibonacci sequence. The collapse at $L - F^{(2)}_n = 0$ is another representation of 
the power-law scaling observed in left panel of Fig. \ref{fig:2ndFibo}, with the constant width of the central peak for various scalings $F^{(2)}_n=322, 521, 843, 1364$ indicative of its stability with respect 
to fluctuations around these lengths. At $m=1$ Fibonacci lengths, further peaks are observed indicated by dashed lines.}
\label{fig:2ndpeaks}
\end{figure}

\begin{figure}[bp!]
\centerline{\includegraphics[width=3.3in]{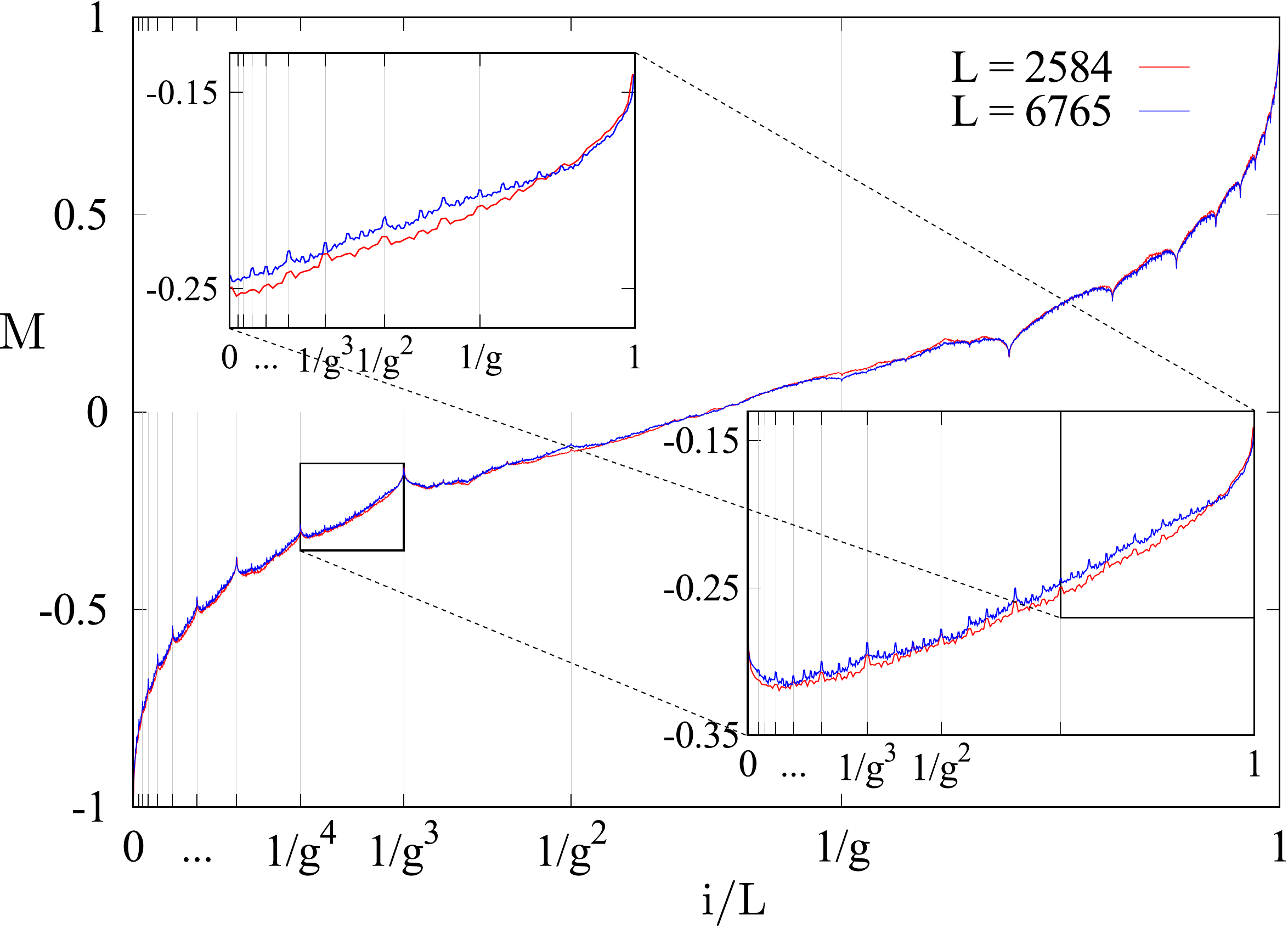}}
\vspace{-0.5cm}
\caption{(Color online) Magnetization profile in the Fibonacci model at $h=0.5$ for Fibonacci chain lengths. Like in the \AAH{} model's result in Fig. 1 of the main text, 
finer and finer details are revealed with peaks located at fractions of the inverse golden ratio. 
Fractal dimension is $D_{\textrm{f}} \approx 1.09$. $600 - 1000$ disorder samples were used.}
\label{fig:Fibo_w_inset}
\end{figure}
\begin{figure}[tp!]
\centerline{\includegraphics[width=3.5in]{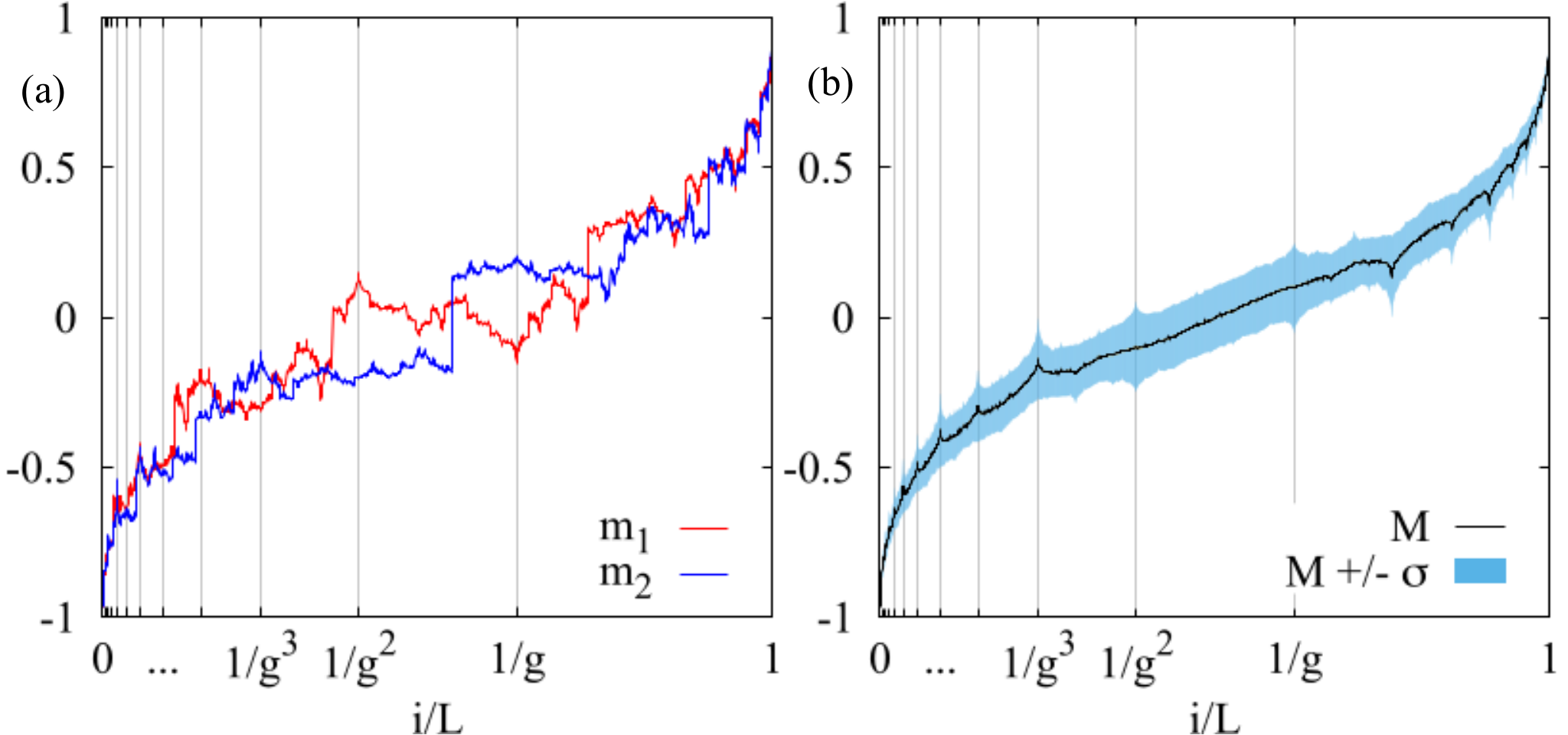}}
\vspace{-0.4cm}
\caption{(Color online) Magnetization profile in the Fibonacci model at $h=0.5$ and Fibonacci chain lengths $L=1597$. 
Left panel:~Profiles $m$ for two different realizations of the disorder (no averaging).
Right panel:~Profile $M=\overline{m}$ obtained upon averaging over $1100$ disorder samples. The error bars on the estimation of $M$ are smaller than the thickness of the line. 
The light blue background is defined by the standard deviation of $m$ from its average value $M$,  
in which profiles are more likely to lie from one realization of the disorder to another.
}
\label{fig:Fibo_Av_vs_NonAv}
\end{figure}
\begin{figure}[bp!]
\centerline{\includegraphics[width=3.3in]{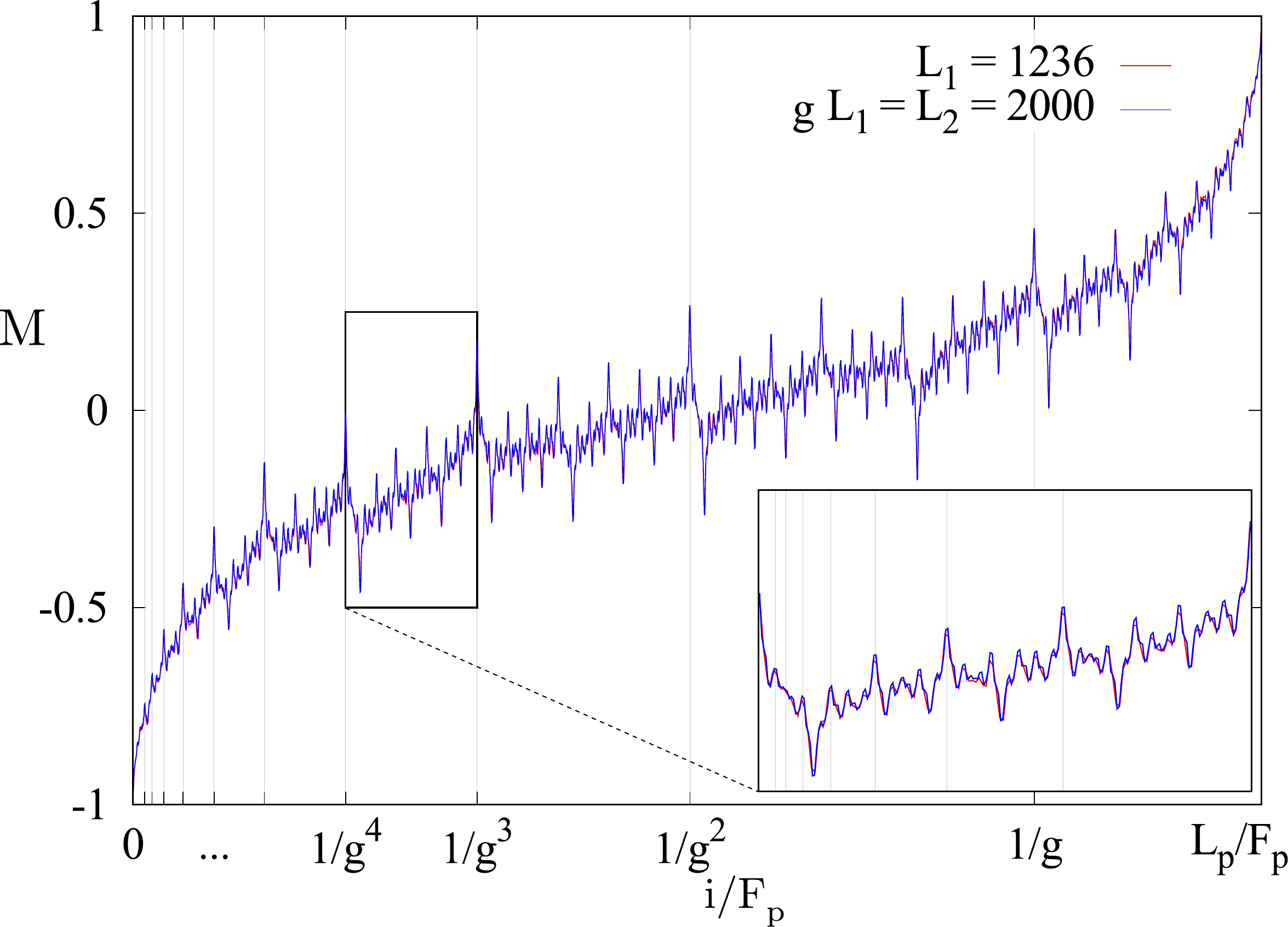}}
\vspace{-0.5cm}
\caption{(Color online) Magnetization profile in the Fibonacci model at $h=0.5$ for non-Fibonacci chain lengths. 
Here the rescaling of the x-axis was done by $F_p$, which is the smallest Fibonacci number larger than $L_p$.  
We find that there is a universality among the magnetization profiles for lengths related by $\tilde{L} = L/g^n$, for integer $n$. 
$1000$ ($L= 2000$) - $2000$ ($L= 1236$) disorder samples were used.
}
\label{fig:Fibo_w_inset2}
\end{figure}
\begin{figure}[tp!]
\centerline{\includegraphics[width=3.3in]{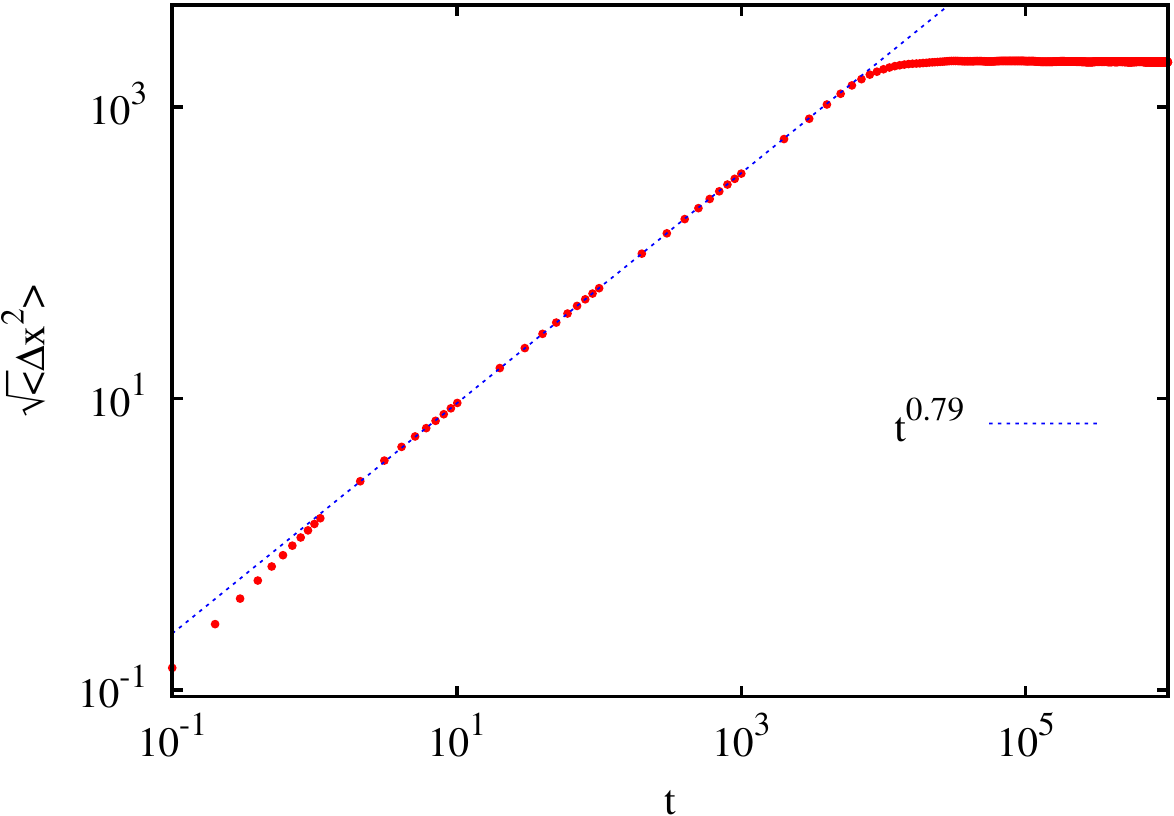}}
\vspace{-0.5cm}
\caption{(Color online) Root mean-squared displacement in wavepacket spreading with Hamiltonian dynamics in Fibonacci Hamiltonian with free boundaries, $L=8001$. 
A superdiffusive scaling is observed (results have been sample-averaged as explained in the text, but results are the same even for single realization i.e. the sequence given by Eq. \eqref{eq: FibSeq}). 
The exponent is consistent with that observed in Ref. \cite{Abe} but not with that obtained whilst employing scaling relation $\beta = 1/(1+ \gamma)$ \cite{ZnidaricVarma} 
(see Fig. 3 of the main text for $\gamma$ results), as also observed in the Aubry-Andr\'{e}-Harper model.
}
\label{fig:FiboClosed}
\end{figure}
\begin{figure*}[!t]
\centerline{\includegraphics[width=6.5in]{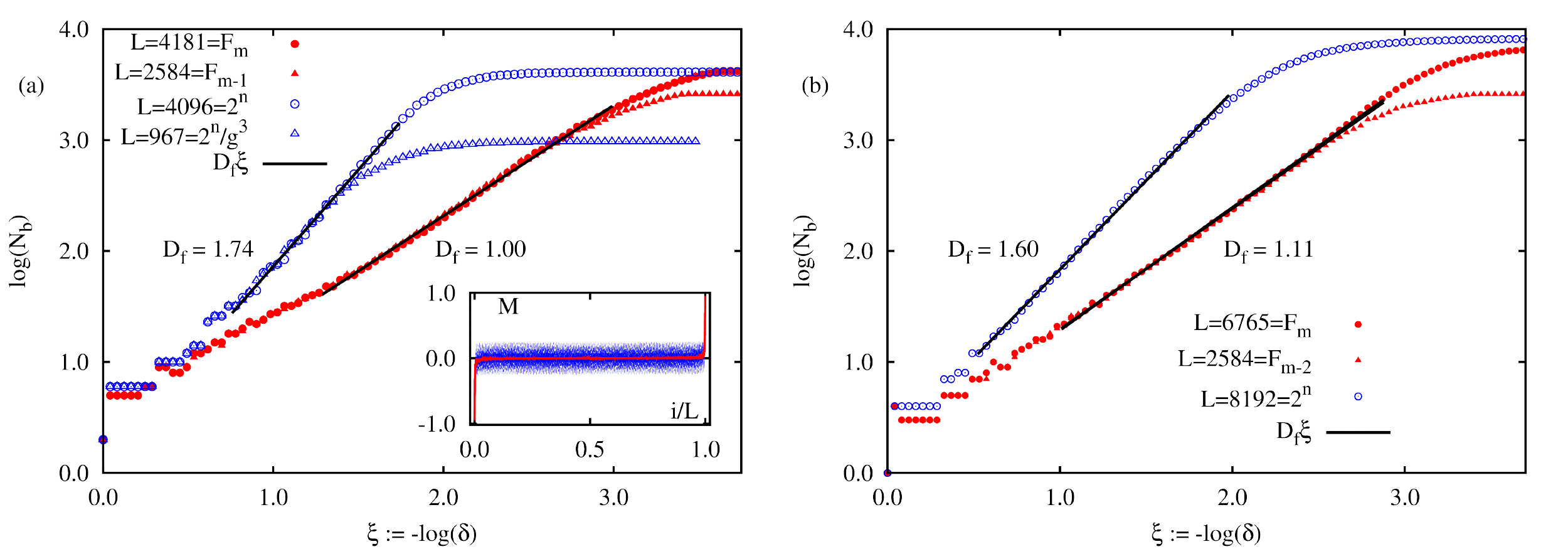}}
\vspace{-0.2cm}
\caption{(Color online) Box counting dimension $D_{\textrm{f}}$ of NESS magnetization profiles in \AAH{} model obtained as linear fit to logarithm of number $N_b$ of boxes required to cover the data to the logarithm of the box size $\delta$. 
Left panel: ballistic phase with $h=1.5$; for $L = F_m$ (red closed symbols) there is no fractality, which emerges when $L \neq F_m$ (blue open symbols). 
Inset shows the magnetization profiles for Fibonacci (red lines) and non-Fibonacci (blue lines) lengths used to compute the $D_{\textrm{f}}$ in the main panel.
Right panel: critical point with $h=2.0$; there is fractality both for $L=F_n$ and for $L \neq F_m$, with the latter having a larger fractal dimension.
%
}
\label{fig:boxcounting}
\end{figure*}
In the main text as well as in the next subsections we present results showing fractality of the magnetization profiles. 
However as we claimed in the abstract this fractality is a generic feature in these systems and must be visible in other observables as well.
We may convince ourselves that this is the case by visualizing the solution of the complex correlation matrix $C$, whose two particular entries are the magnetization (diagonal) and the current (super/sub diagonal).

In Fig.~\ref{fig:C_n610} we show a density plot of the correlation matrix for the $L=610$ Fibonacci length \AAH{} model at $h=h_c=2$.
In particular if we segment this matrix into smaller (Fibonacci) lengths, we will see the larger structure grossly replicated at the smaller lengths too. This picture also suggests that, if one drives the system at sites with Fibonacci number index, where excitations are small, the fluctuations and fractal dimensions will indeed be smaller as demonstrated for magnetization profiles. 
Other observables too must display the feature of fractality.

In Fig.~\ref{fig:AAHsigmaFib} we show fractal dependence of profile fluctuations in the \AAH{} model (taking an ensemble with random phases).  

\subsection{Nonprimary resonances and off-criticality}
In Fig. 3 of main text we presented scaling results of the NESS current at criticality. We observed that a whole series of ``satellite'' resonances appeared that lay between the Fibonacci length and the generic length systems.
We demonstrate here that these all fall within a well-defined sequence.

Replacing the Fibonacci number sequence generating rule $F_n = F_{n-1} + F_{n-2}$ by $F^{(m)}_n = F_{n-1} + F_{n-m-1}$ we obtain an $m-$Fibonacci sequence which are also related, as in the Fibonacci rule, by
\begin{equation}
 \label{eq: mFibo}
 F^{(m)}_n = F^{(m)}_{n-1} + F^{(m)}_{n-2},
\end{equation}
with the ratio between two large successive integers approaching the golden ratio. For $m=2$, these secondary resonances were labelled in the inset of Fig. 3 of main text: $L=322, 521, 843, 1364$. Although these resonances, at first 
glance, look like noise as displayed in the inset we demonstrate that these too display a clean scaling behaviour: there's method in its madness. 
This is shown in the left panel of Fig. \ref{fig:2ndFibo} where a nice scaling to subdiffusive behaviour fits the 
data. The exponent $\gamma$ is very close to that obtained for the Fibonacci lengths $m=1$, which was plotted in the main text. 

The stability of these resonances can also be visualized more conspicuously by plotting the dependence of the scaled NESS data for various fixed $L$ as a function of its deviation from the secondary Fibonacci sequence.
In Fig. \ref{fig:2ndpeaks} the collapse is indicated: the central peak has a common height for the four scalings, as well as a constant width. 
The first point is indicative of the power-law scaling observed in left panel of Fig. \ref{fig:2ndFibo}, with the power being slighter larger here due to smaller system sizes. 
The second point reflects the stability of the scaling as a function of fluctuations around these secondary Fibonacci lengths. Certain smaller peaks are observed also at the usual Fibonacci numbers, indicated 
by vertical dashed lines, suggesting a whole heirarchy of resonances.

Next we show that these power-law scalings disappear even if marginally away from $h=h_c = 2$, where $h_c$ is the critical point for the \textit{closed} system. 
Therefore $h_c$ remains to be the critical point even for the Lindblad driven system. In the right panel of Fig. \ref{fig:2ndFibo} we show, for Fibonacci lengths, the scaling of the NESS current just left of and 
just right of the transition point. In the former case, the NESS current saturates implying ballistic transport in the metallic phase; whereas in the latter case the current decays exponentially, indicative 
of single-particle localization. 

\subsection{Fibonacci model}
The Fibonacci sequence required for the Fibonacci model may be constructed from two symbols ${F,S}$ by the substitution rule
$
 \left(
 \begin{array}{c}
 F \\ S
 \end{array}
 \right)
 \rightarrow
 \left(
 \begin{array}{cc}
 1 & 1 \\ 1 & 0
 \end{array} \right)
  \left( \begin{array}{c} F \\ S \end{array} \right).
$
The transformation matrix has the eigenvalues $g, 1/g$ where $g$ is the golden ratio. 
Repeated application of the above rule [with summations $+$ arising during matrix multiplication taken to mean concatenation i.e. $F+S \rightarrow FS$] 
gives the series of Fibonacci sequences: 
\begin{equation}
\label{eq: FibSeq}
 \{F, FS, FSF, FSFFS, FSFFSFSF, \ldots \}.
\end{equation} 
Note that the length of each sequence is a Fibonacci number: $1,2,3,5,8 \ldots$, and that, by construction, any sequence always starts with the sequence of any smaller one.
A given cut of length $L$ of any (sufficiently long) Fibonacci sequence determines one sample of the quasiperiodic chain of length $L$, where 
the quasidisorder potential $h_k$ on site $k$ takes the value $\pm h$ depending on whether the symbol on that site is 
$S$ or $F$ respectively; note that the full long sequence is of Fibonacci length but that need not be true for $L$, the system size under study. We choose $h=0.5$ in the work.

The NESS is computed for this $L-$segment, and the procedure is repeated over a different segment of this 
sequence and the results are then averaged over (see Fig.~\ref{fig:Fibo_Av_vs_NonAv}).

In the main text we presented results for current scaling in the superdiffusive regime of the Fibonacci model, and found that Fibonacci lengths and generic lengths harbour different rates of transport, 
as evidenced by the scaling of NESS current with system size. 
Here we present evidence that, like the critical \AAH{} model, the magnetization profile shows features of fractality. In Fig.~\ref{fig:Fibo_w_inset} we show the disorder averaged magnetization profiles 
for two Fibonacci length systems. It is seen that upon zooming in, there are regular finer peaks and structures located at certain powers of $1/g$ as shown by the thin vertical lines. 
A similar structure is observed for non-Fibonacci chain lengths in Fig.~\ref{fig:Fibo_w_inset2}, where the universality amongst different non-Fibonacci length chains is made apparent. 
Note that the $x=1/g^k$ positions, for integer $k$, highlighted after rescaling, correspond, before the rescaling, 
to the Fibonacci number lattice indices and to their possible combinations: $i=F_n$ (main), $i=F_n\pm F_m$ (inset), etc.. 
Observe how the overall shape of the 
profile has a curvature of opposite sign compared to that of Fig. 1 of the main text. This just reflects the difference in dynamics between the two cases: 
here we are in the superdiffusive regime (because we chose a small $h=0.5$), while the critical \AAH{} is subdiffusive. 
Apart from these fine fractal features we have found that the gross shape is captured by the Beta function (not shown), see e.g. Ref.~\cite{ZnidaricVarma} for formulas.

In addition to nonequilibrium dynamics, we also performed wavepacket spreading computations on the Fibonacci lattice with $h=0.5$, akin to the computations of Ref. \cite{Abe} (however there the authors considered a symmetrized version of the 
Fibonacci potential). 
We show in Fig. \ref{fig:FiboClosed} wavepacket spreading dynamics and find a superdiffusive spreading with an exponent $\beta = 0.79$. This is consistent with that seen in Ref. \cite{Abe}; however, as observed in 
the main text for the Aubry-Andr\'{e}-Harper Hamiltonian, it is not consistent with the relation $\beta = 1/(1+\gamma)$ which is otherwise deemed to be valid \cite{ZnidaricVarma}; the $\gamma$ values for 
nonequilibrium transport have been shown in Fig. 3 of the main text. The results are unchanged whether we take the sequence Eq. \eqref{eq: FibSeq} or the sequences are position-averaged (which procedure we explained earlier).
\subsection{Box counting dimension}

Box counting is a simple procedure to assess fractality of a given spatial pattern. Summarily put, it counts the power with which the number of boxes $N_b$ required to cover the data points 
increases as the size of the box $\delta$ decreases i.e. $N_b = \delta^{-D_{\textrm{f}}}$, with $0 < \delta < 1$ due to appropriate rescaling of data.
Numerically the fractal dimension $D_{\textrm{f}}$ may be computed by finding the linear slope of $\log{N_b}$ versus $\log{\delta}$ i.e. $D_{\textrm{f}} = -\cfrac{\log{N_b}}{\log{\delta}}$.
A simple one-dimensional pattern formed from simple lines and wiggles that has no fine structure will have a fractal dimension $D_{\textrm{f}} = 1$; similarly a two-dimensional pattern will have $D_{\textrm{f}} = 2$.
Fractality of magnetization profiles will be characterized by $1 < D_{\textrm{f}} < 2$.

In Fig.~\ref{fig:boxcounting} we show the evaluation of the box-counting dimension in the \AAH{} model in the ballistic phase and at the critical point $h = 2.0$. 
In each case we compute $D_{\textrm{f}}$ when the chain length is a Fibonacci number or not. The plot of $\log{N_b}$ versus $\log{1/\delta}$ in each case yields a clear linear regime, whose slope 
gives us the box-counting fractal dimension $D_{\textrm{f}}$.

In the ballistic phase ($h=1.5$) we see that for a Fibonacci length chain a good fit with $D_{\textrm{f}} = 1.00$ is obtained; this is also clear from the inset where the red full line ($L=4181$, Fibonacci length) 
shows a smooth ballistic magnetization profile, making it immediately evident from a simple visual inspection that there is no fractal structure. 
However the blue dashed line ($L=4096$, non-Fibonacci length) in the inset shows oscillations around the red full line; its fractal dimension is therefore substantially larger (almost reaching 2) as seen from the fit in the 
main panel. 900 phase averages were performed for both cases.

At criticality (right panel of Fig.~\ref{fig:boxcounting}) the qualitative picture remains the same i.e., away from Fibonacci lengths the fractal dimension increases.
However here both Fibonacci and non-Fibonacci length lattices bear magnetization profiles that are fractal, with $D_{\textrm{f}} \approx 1.11$ for the former case, not too different from (but still unequal to) 
the lattice dimensionality. A similar $D_{\textrm{f}} \approx 1.09$ is found for the Fibonacci model (results not shown).

\end{document}